\documentclass[preprintnumbers,article,amsmath,amssymb,floatfix,10pt,prd,onecolumn,
superscriptaddress,nofootinbib]{revtex4}
\usepackage{bbm}
\usepackage{amsfonts}
\usepackage{mathrsfs}
\usepackage{latexsym}

\usepackage{epsfig}
\usepackage{graphicx}
\usepackage{epstopdf}
\usepackage{placeins}
\usepackage{float}

\usepackage{amssymb}
\usepackage{amsmath}

\usepackage{dcolumn}
\usepackage{bm}
\usepackage{color}
\usepackage{comment}
\usepackage{xcolor}

\begin{document}
\title{\bf Relativistic Krori-Barua Compact Stars in $f(R,T)$ Gravity}
\author{M. Farasat Shamir}
\email{farasat.shamir@nu.edu.pk}\affiliation{National University of Computer and
Emerging Sciences,\\ Lahore Campus, Pakistan.}
\author{Zoya Asghar}
\email{zoyaasghar16411@gmail.com}\affiliation{National University of Computer and
Emerging Sciences,\\ Lahore Campus, Pakistan.}
\author{Adnan Malik}
\email{adnan.malik@skt.umt.edu.pk}\email{adnanmalik_chheena@yahoo.com}\affiliation{Department of Mathematics, University of Management and Technology,\\ Sialkot Campus, Pakistan.}

\begin{abstract}
This work aims to investigate the behaviour of compact stars in the background of $f(R, T)$ theory of gravity. For current work, we consider the Krori-Barua metric potential i.e., $\nu(r)= Br^2+C$ and $\lambda(r)= Ar^2,$ where, $A, B$ and $C$ are constants. We use matching conditions of spherically symmetric space-time with Schwarzschild solution as an exterior geometry and examine the physical behaviour of stellar structure by assuming the exponential type $f(R, T)$ gravity model. In the present analysis, we discuss the graphical behaviour of energy density, radial pressure, tangential pressure, equation of state parameters, anisotropy and stability analysis respectively.  Furthermore, an equilibrium condition can be visualized through the modified Tolman-Oppenheimer-Volkov equation. Some extra features of compact stars i.e. mass-radius function, compactness factor and surface redshift have also been investigated. Conclusively, all the results in current study validate the existence of compact stars under exponential $f(R, T)$ gravity model.
\end{abstract}
\maketitle
\section{Introduction}
The spatial behaviour of accelerated expansion of the universe is the most attractive topic of discussion among astrophysicists, and this expansion of the universe relies on dark matter $(DM),$ ordinary matter $(OM),$ and dark energy $(DE)$ \cite{Ast,Eis,Rie,Sper}. Therefore, General relativity $(GR)$ provides the most satisfactory results and develops a basic understanding of gravitational theories. But $GR$ alone does not give more appropriate outcomes to determine the mysterious nature of $DE.$ Due to the limitation of $GR,$ the modified theories have become more popular among cosmologists. Different modified theories are available in the literature such as, $f(R),$ $f(G),$ $f(T),$ $f(Q),$ $f(R,T)$, $f(R,G)$, $f(R, \phi)$ and $f(R, \phi, X)$ theories of gravity \cite{Ahm}-\cite{ad15}. Interestingly in recent times, astrophysicists intend to investigate the stability and collapsing aspect of stellar structures with the help of modified theories of gravity. Cognola et al. \cite{Cog} presented the exponential type gravity models to study the physical behaviour of stellar structures. Naz et al. \cite{Naz} investigated the existence of a new classification of embedded class-I solutions of compact stars, by using the Karmarkar condition in $f(R)$ gravity background. Later, Shamir et al. \cite{Naz1} discussed the physical analysis of charged anisotropic Bardeen spheres in the $f(R)$ theory of gravity with the Krori-Barua metric. Harko \cite{Har} proposed the $f(R,T)$ theory of gravity, which is a combination of the Ricci scalar and trace of the energy-momentum tensor.
Moreas et al. \cite{Mor} studied the hydrostatic equilibrium configuration of neutron stars and strange stars in the background of the $f(R, T)$ theory of gravity. Das et al. \cite{Das} generated a set of solutions describing the interior of a compact star under the $f(R, T)$ theory of gravity, which admits conformal motion. Later, Waheed et al. \cite{Wah} explored the existence of a new family of compact star solutions by adopting the Karmarkar as well as Pandey-Sharma condition in the background of $f(R, T)$ modified gravitational framework. Ilyas \cite{Ily} explored and analyzed a set of solutions describing the interior structure of relativistic compact stellar structures with the variable cosmological constant in the $f(R, T)$ theory of gravity. Zubair et al. \cite{Zub} constructed an anisotropic solution for spherically symmetric space-time that satisfies Karmarkar condition in the context of the $f(R,T)$ theory of gravity. Furthermore, Pretel et al. \cite{Pre} investigated the equilibrium and radial stability of spherically symmetric relativistic stars, considering a polytropic equation of state, within the framework of $f(R, T)$ gravity with a conservative energy-momentum tensor. Islam et al. \cite{aa1} presented the interior solutions of distributions of magnetized fluid inside a sphere in $f(R,T)$ gravity, in which the magnetized sphere is embedded in an exterior Reissner–Nordström metric. Shamir and Waseem \cite{aa2} analyzed the nature of anisotropic spherically symmetric relativistic star models in the framework of $f(R, T)$ theory of gravity by considering the MIT bag model equation of state. Lobato et al. \cite{aa3} investigated the neutron stars in $f(R,T)$ gravity using realistic equations of state in the light of massive pulsars and GW170817.  \\

Investigating the behaviour of stellar structures by using the Krori-Barua metric has been a very fascinating topic among researchers. Bhar \cite{KB1} provided a new model of a hybrid star with strange quark matter along with normal baryonic matter in the framework of Krori and Barua ansatz. Rahaman et al. \cite{KB2} explored the possibility of applying the Krori and Barua model to describe ultra-compact objects like strange stars and the consequences of a mathematical description to model strange stars have been analyzed. Biswas et al. \cite{KB3} investigated the anisotropic charged strange stars in Krori-Barua spacetime under $f(R,T)$ theory of gravity. Sharif and Waseem \cite{KB4} investigated the behaviour of anisotropic compact stars in the background of $R + \alpha R_{\mu \nu} T^{\mu \nu}$ gravity model and used Krori-Barua metric solutions, where constants are calculated using masses and radii of compact stars like Her X-1, SAX J 1808.4–3658, and 4U1820–30. Abbas et al. \cite{KB5} studied the possibility of forming anisotropic compact stars in modified Gauss-Bonnet, namely called as $f(G)$ theory of gravity which is one of the strong candidates, responsible for the accelerated expansion of the universe and used an analytical solution of Krori and Barua metric to the Einstein field equations with an anisotropic form of matter and power law model of $f(G)$ gravity. Shamir and Malik \cite{KB6} discussed the behaviour of anisotropic compact stars using Krori-Barua metric in $f(R, \phi)$ theory of gravity, where $R$ and $\phi$ denote the Ricci scalar and scalar field respectively. Malik and his collaborators \cite{ad5,ad10} investigated the behaviour of charged compact stars in the modified $f(R)$ theories of gravity and assumed the Krori- Barua space-time to describe the geometry of the inner space, which is a singularity-free solution for a charged fluid sphere in general relativity. Bhar \cite{KB7} considered the spherically symmetric spacetime along with anisotropic fluid distribution in the presence of an electric field in $f(R, T)$ theory of gravity admitting the Chaplygin equation of state. Hossein et al. \cite{Hos} discussed the possibility of forming anisotropic compact stars from this cosmological constant as one of the competent candidates of dark energy by taking the analytical solution of the Krori and Barua metric. Later, Rehman et al. \cite{Reh} explored the possibility of applying the Krori and Barua model to describe ultra-compact objects like strange stars and some bounds on the model parameters have been obtained for making it a viable model for strange stars.\\

After being inspired by the preceding literature, we further extend the idea of Shamir et al. \cite{Naz1} in modified $f(R, T)$ gravity and investigate the physical behaviour of different compact stars by employing Krori-Barua spacetime. For our current study, we choose the exponential type $f(R,T)$ gravity model, i.e., $f(R,T)=R+\alpha(e^{-\beta R}-1)+\lambda T$ \cite{Cog}, which includes the Lagrangian matter $\mathcal{L}_{m}=-\frac{1}{3}(p_{r}+2 p_{t})$ \cite{Har}. The arrangement of our manuscript is as follows: Section $II$ is devoted to the mathematical formulation of $f(R, T)$ gravity in the framework of anisotropic distributions. Section $III$ deals with the $f(R, T)$ gravity model and matching conditions for determining the unknown constants by comparing the intrinsic metric to Schwarzschild's exterior metric. Section $IV$ is used to describe the physical analysis, such as anisotropic parameter, energy conditions, equation of state parameters, stability analysis, Tolman-Oppenheimer-Volkoff equation, mass-radius relationship, compactness factor, and surface redshift analysis. In the last section, we finalized the conclusive remarks.
\section{Modified $f(R,T)$ Field Equations}
\begin{figure}[h!]
\begin{tabular}{cccc}
\epsfig{file=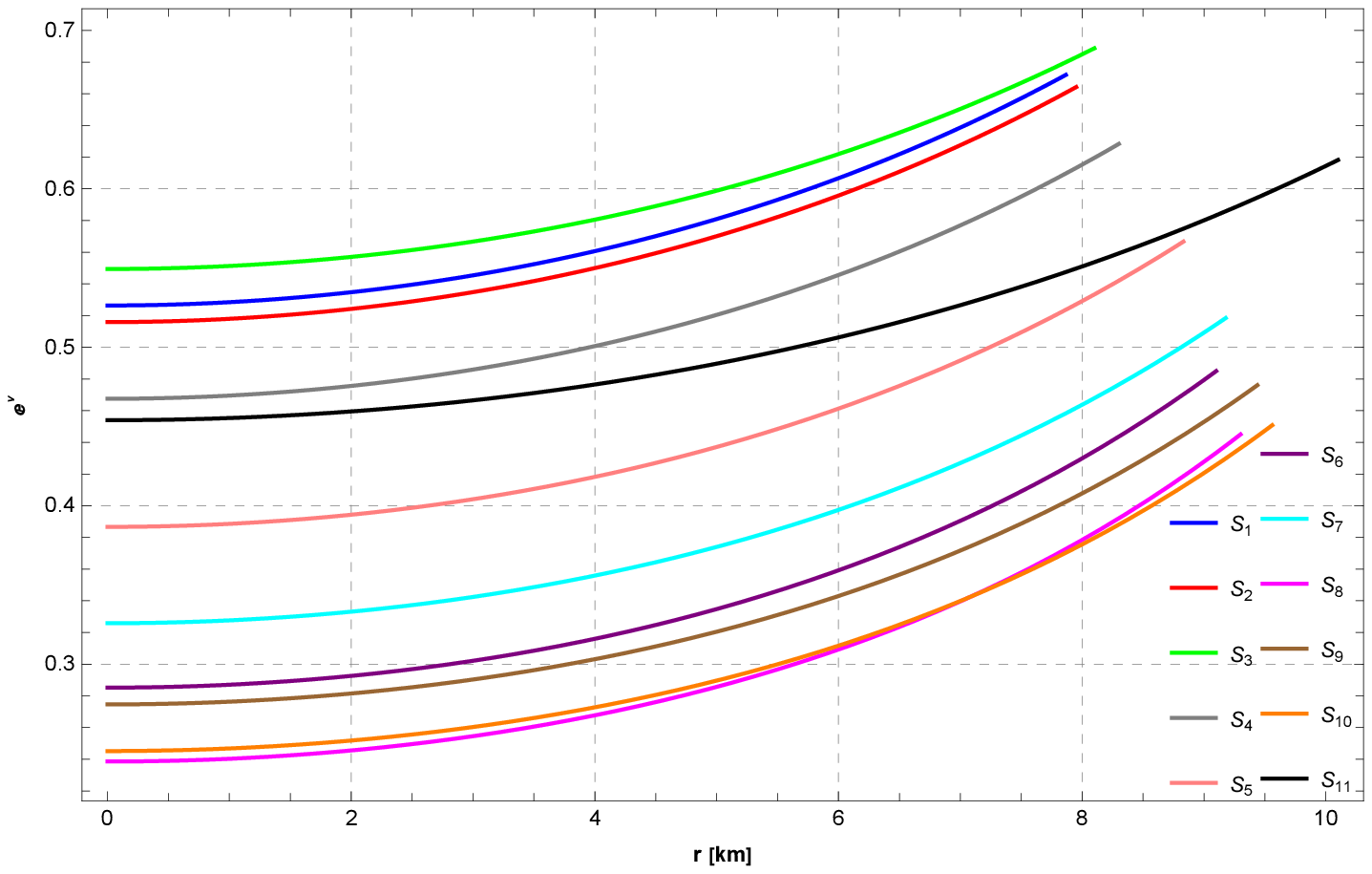,width=0.33\linewidth} &
\epsfig{file=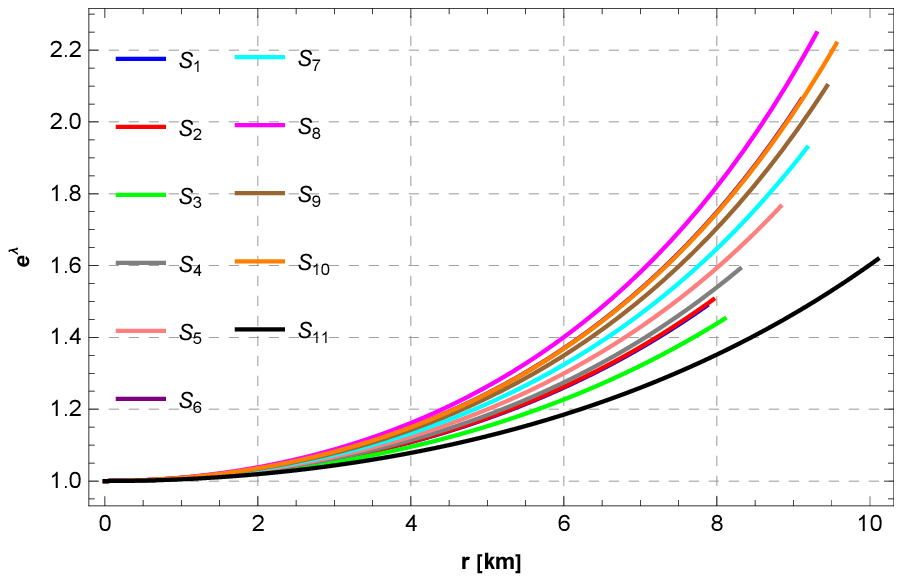,width=0.33\linewidth} &
\end{tabular}
\caption{{Graphical behavior of metric potentials}.}
\label{Fig:1}
\end{figure}

The Einstein-Hilbert action in modified $f(R,T)$ gravity is given as
\begin{equation}\label{0}
S= \int\sqrt{-g}\Big(\frac{1}{2\kappa}f(R,T)+\mathcal{L}_{m}\Big)d^4x,
\end{equation}
where,
\begin{itemize}
    \item $f$ is a function that depend on $R$ and $T$,
    \item $\mathcal{L}_{m}$ is Lagrangian matter,
    \item $\kappa=\frac{8\pi G}{c^{4}}$ denotes coupling constant,
    \item $g$ represents the determinant of metric $g_{\eta\zeta}$.
\end{itemize}
By varying the action (\ref{0}) with respect to metric tensor, we get modified $f(R,T)$ gravity field equation as
\begin{equation}\label{1}
f_{R}(R,T) R_{\eta\zeta}-\frac{1}{2} f(R,T)g_{\eta\zeta}+(g_{\eta\zeta} \Box-\nabla_\eta\nabla_\zeta)f_R(R,T)= 8\pi T_{\eta\zeta}-f_T(R,T)T_{\eta\zeta}-f_T (R,T) \Theta_{\eta\zeta},
\end{equation}
where, $\Box\equiv\nabla_{\eta}\nabla^{\zeta}$ denotes the D'Alembertian symbol and $\nabla_\eta$ symbolizes covariant derivative, whereas, $f_R(R,T)=\frac{\partial f(R,T)}{\partial R}$ and $f_T(R,T)=\frac{\partial f(R,T)}{\partial T}$ . Furthermore, $T_{\eta\zeta}$ indicates the anisotropic energy momentum tensor is defined as
\begin{equation}\label{2}
T_{\eta \zeta}=(\rho+p_{t})u_{\eta}u_{ \zeta}-p_{t}g_{\eta \zeta}+(p_{r}-p_{t})\mathcal{X}_{\eta}\mathcal{X}_{\zeta}.
\end{equation}
Here, $\mathcal{X}_{\eta}$ and $\mathcal{X}_{\zeta}$ denote the four velocity vector with $u^{\eta}u_{\eta}=-\mathcal{X}^{\eta}\mathcal{X}_{\eta}=1$, whereas, $\rho$, $p_r$ and $p_t$ represent energy density and radial pressure and tangential pressure respectively.
The covariant divergence becomes \cite{Barri} presents
\begin{equation}\label{3}
\nabla^\eta T_{\eta\zeta} = \frac{f_{T}(R,T)}{8\pi-f_{T}(R,T)}\Big((T_{\eta\zeta}+\Theta_{\eta\zeta})\nabla^\eta \ln f_{T}(R,T)+\nabla^\eta \Theta_{\eta\zeta}-\frac{1}{2}g_{\eta\zeta}\nabla^\eta T \Big),
\end{equation}
The non-zero right hand side of Eq. (\ref{3}) indicates that the $f(R,T)$ theory is covariantly divergent. The modified field equations (\ref{1}) can be expressed as
\begin{equation}\label{4}
G_{\eta\zeta}=\frac{1}{f_{R}(R,T)}\Big[8\pi T_{\eta\zeta}+\frac{1}{2}(f(R,T)-R f_{R}(R,T))g_{\eta\zeta}-f_{T}(R,T)(T_{\eta\zeta}+\Theta_{\eta\zeta})-(g_{\eta\zeta}\Box- \nabla_\eta\nabla_\zeta)f_{R}(R,T)\big].
\end{equation}
Here, $\Theta_{\eta\zeta}$ is known as a scalar expansion and defined as $\Theta_{\eta\zeta}=-2 T_{\eta\zeta}-P g_{\eta\zeta}$. We take $\mathcal{L}_{m}=-\mathcal{P},$ where $\mathcal{P}=\frac{1}{3}(p_{r}+2 p_{t})$ \cite{Har}. The spherically symmetric spacetime is defined as
\begin{equation}\label{5}
ds^{2}=e^{\nu}{dt}^{2}-e^{\lambda}{dr}^{2}-r^{2}({d\theta}^{2}+sin^{2}\theta{d\phi}^{2}),
\end{equation}
where, $\lambda$ and $\nu$ are functions of $r$. Moreover, we consider the Krori and Barua metric potential i.e., $\nu(r)= Br^2+C$ and $\lambda(r)= Ar^2,$ where, $A, B$ and $C$ are the constants. The anisotropic matter configuration must satisfy the conditions that $e^{\lambda}|_{r=0}=1$ and $e^{\nu}|_{r=0}=e^C,$ moreover, $(e^{\lambda})^{'} = 2Ae^{Ar^2} r$ and $(e^{\nu})^{'} = 2Be^{C+Br^{2}} r.$ As a result, the metric potential determined here assures that the potentials are well-behaved, nonsingular, and steady in the core of the compact star. The fundamental advantage of employing this approach is that it yields a singularity-free model. Fig. \ref{Fig:1} exhibits that both metric potentials are minimum at the center and maximum at the surface boundary.
\FloatBarrier

\section{$f(R,T)$ gravity model and Matching Conditions}
Here, we choose an exponential model of the $f(R,T)$ gravity to examine the stability and consistency of the stellar system. This model helps us for obtaining a comprehensive understanding of the physical structure of the stellar star. The $f(R,T)$ gravity model \cite{Cog} is defined as
\begin{equation}\label{6}
f(R,T)=R+\alpha(e^{-\beta R}-1)+\gamma T,
\end{equation}
where, $\alpha,$ $\beta$ and $\gamma$ are constant. This model is preferred because it exhibits exponential growth for the early phases of cosmic evolution. Cognola et al. \cite{Cog} studied exponential-type models to determine the physical attributes of compact stars and claimed that the model satisfies every local test involving the consistency of spherical object problems and non-violation of Newtonian's law. Substituting all the values in Eq. (\ref{1}), we get the following set of field equations as
\begin{eqnarray}\label{7}
\nonumber &&
\rho = \mathcal{Z}_{1} \Big(\frac{1}{(\gamma +8 \pi ) r^2}\Big(e^{-A r^2} \Big(4 \gamma  \mathcal{Z}_{3} \left(-A B r^4+8 A r^2+4 e^{A r^2}+B^2 r^4+3 B r^2-4\right)-2 \gamma  r \left(\mathcal{Z}_{4} \left(5 A r^2-3 B r^2+14\right)+7 r \mathcal{Z}_{5}\right)~~~
\\&&+96 \pi  \left(\mathcal{Z}_{3} \left(2 A r^2+e^{A r^2}-1\right)-r \left(A r^2 \mathcal{Z}_{4}+r \mathcal{Z}_{5}+2 \mathcal{Z}_{4}\right)\right)\Big)\Big)-6 \alpha  \beta  \mathcal{Z}_{2} e^{-\beta  R(r)}+\alpha  \left(6-6 e^{\beta  (-\mathcal{Z}_{2})}\right)\Big),
\end{eqnarray}
\begin{eqnarray}\label{8}
\nonumber &&
p_{r} = \mathcal{Z}_{1} \Big(\frac{1}{(\gamma +8 \pi ) r^2}\Big(2 e^{-A r^2} \Big(\gamma  \mathcal{Z}_{3} \left(2 A B r^4+8 A r^2-8 e^{A r^2}-2 B^2 r^4+18 B r^2+8\right)-48 \pi  \mathcal{Z}_{3} \left(e^{A r^2}-2 B r^2-1\right)+~~~
\\&& \gamma  r \left(\mathcal{Z}_{4} \left(17 A r^2+9 B r^2+14\right)-5 r \mathcal{Z}_{5}\right)+48 \pi  r \mathcal{Z}_{4} \left(2 A r^2+B r^2+2\right)\Big)\Big)+6 \alpha  \beta  \mathcal{Z}_{2} e^{-\beta  \mathcal{Z}_{2}}+6 \alpha  \left(e^{\beta  (-\mathcal{Z}_{2})}-1\right)\Big),
\end{eqnarray}
\begin{eqnarray}\label{9}
\nonumber &&
p_t = \mathcal{Z}_{1}\Big(\frac{1}{(\gamma +8\pi) r^2}\Big(e^{-A r^2} \Big(-2 A r^2 \left(2 \gamma \mathcal{Z}_{3} \left(5 B r^2+2\right)+48 \pi \left(B r^2 \mathcal{Z}_{3}-r \mathcal{Z}_{4}+\mathcal{Z}_{3}\right)-5 \gamma r \mathcal{Z}_{4}\right)-6 \alpha \gamma r^2 e^{A r^2}-48 \pi \alpha r^2 ~~~
\\ \nonumber && e^{A r^2}+8 \gamma \mathcal{Z}_{3} e^{A r^2}+4 B^2 (5 \gamma +24 \pi ) r^4 \mathcal{Z}_{3}+6 B (3 \gamma +16 \pi ) r^2 (r \mathcal{Z}_{4}+2 \mathcal{Z}_{3})+14 \gamma  r^2 \mathcal{Z}_{5}+96 \pi r^2 \mathcal{Z}_{5}+4 \gamma r \mathcal{Z}_{4}+96 \pi r \mathcal{Z}_{4}
\\&&-8 \gamma \mathcal{Z}_{3}\Big)\Big)
 +6 \alpha e^{\beta  (-\mathcal{Z}_{2})} (\beta \mathcal{Z}_{2}+1)\Big).
\end{eqnarray}
where,
\\
$\mathcal{Z}_{1}=\frac{1}{24 (\lambda+4 \pi )},$ ~~~~~~~~~$\mathcal{Z}_{2}=-\frac{2 e^{-A r^2} \left(A r^2 \left(B r^2+2\right)+e^{A r^2}-B^2 r^4-3 B r^2-1\right)}{r^2},$~~~~~~~~~$\mathcal{Z}_{3}=(1-\alpha \beta e^{-\beta \mathcal{Z}_{2}}),$ \\
\\
~~~~~~~~~~~~~~~~~~~~~~~~~~~~~$\mathcal{Z}_{4}=\frac{\partial \mathcal{Z}_{3}}{\partial r},$~~~~~~~~~$Z_5=\frac{\partial Z_{4}^2}{\partial r^2}.$\\
\\
\\

The Schwarzschild solution is considered the most appropriate choice for choosing among the numerous matching constraints while investigating compact objects in the background of $GR$. The Jebsen-Birkhoff theorem states that the solution of the Einstein field equations $(EFE)$ must be asymptotically flat and static for spherically symmetric spacetime. When we encounter modified $TOV$ equations with zero pressure and energy density, the exterior geometry solution may differ from the Schwarzschild solution in modified gravitational theories \cite{Tol,opp1}. However, in modified $f(R,T)$ gravity, the solution of Schwarzschild can be satisfied by preferring a realistic $f(R,T)$  gravity model for non-zero density and pressure. Due to this fact, the Birkhoff theorem violates the modified theories of gravity \cite{Far}. Many researchers have investigated matching conditions and provided remarkable results about the Schwarzchild solution \cite{Ast,Coo,Gan,Mome}. Now, to solve the $EFE$ with the restricted condition at $r = R,$ the pressure $p_{r}(r = R) = 0,$  we compare our interior solution to the outer Schwarzschild solution, which is defined by
\begin{equation}\label{10}
ds^{2}=\big(1-\frac{2M}{r}\big){dt}^{2}-\big(1-\frac{2M}{r}\big)^{-1}{dr}^{2}-r^{2}({d\theta}^{2}+sin^{2}\theta{d\phi}^{2}),
\end{equation}
where, $M$ symbolizes the mass of stellar stars. The continuity of the metric functions of Eq. (\ref{4}) at the $r = R$ leads to the following form
\begin{equation}\label{11}
  {g_{rr}}^{+} = {g_{rr}}^{-},~~~~~~
  {g_{tt}}^{+} = {g_{tt}}^{-},~~~~~~
\frac{\partial{g_{tt}}^{+}}{\partial r} = \frac{\partial{g_{tt}}^{-}}{\partial r}.
\end{equation}
Here, $(+)$ denotes the exterior geometry and $(-)$ indicates the interior. Further, we use Eqs. (\ref{4}), (\ref{10}) and (\ref{11}) to get the values of $A,$ $B,$ and $C,$ given as
\begin{equation}\label{12}
A=\frac{1}{R^{2}}\ln[\frac{R}{R-2M}],~~~~
B=\frac{M}{R^{2}(R-2M)},~~~~
C=\ln[\frac{R-2M}{R e^{BR^{2}}}].
\end{equation}
Table \ref{tab1} provides the constants $A, B,$ and $C$ regarding the mass and radius of the chosen compact objects.
Moreover, it is necessary to mention that the following conditions must be fulfilled for the appropriate behavior of compact objects:
\begin{itemize}
  \item The plots of $\rho,$ $p_{r}$ and $p_{t}$ have to be finite and maximum at the center.
  \item The graphs of $\frac{dp_{r}}{dr},$ $\frac{dp_{t}}{dr}$ and $\frac{d\rho}{dr}$ should be negative.
  \item The energy conditions must be satisfied.
  \item It is necessary the radial and tangential equation of state must lie between zero and 1, i.e., $0 < w_{r},$ $w_{t} < 1.$
  \item The radial and transverse sound speed satisfy the range i.e., $0 < v_{r}^{2}, v_{t}^{2} < 1.$
  \item All forces should satisfy the equilibrium condition.
\end{itemize}
\begin{table}[ht]
\centering
\caption{ Unknown parameters of the stellar objects for $\alpha=-2.5$ and $\gamma=0.4$}.
\begin{tabular}{|c|c|c|c|c|c|c|}
  \hline
  ~
  % after \\: \hline or \cline{col1-col2} \cline{col3-col4} ...
  $\textbf{Star Model}$ & $M~(M_{\Theta})$ & $R~(km)$ & $A~(km)$ & $B~(km)$ & $C~(km)$ & $\beta$ \\
  \hline
  $\textbf{4U~1538-52}$ ~
     $(S_{1})$  & ~0.87 $\pm$ 0.07~\cite{Gang} & ~7.866 $\pm$ 0.21 & 0.00642603 & 0.00394556 & -0.641731 & 1.5 \\
  \hline
  $\textbf{SAX~J1808.4-3658}$ ~
     $(S_{2})$  & ~0.9 $\pm$ 0.3~\cite{Gang} & ~7.951 $\pm$ 1.0 & 0.00647132 & 0.00399782 & -0.661843 & 1.51 \\
  \hline
  $\textbf{Her~X-1}$ ~
     $(S_{3})$ & ~0.85$\pm$0.15~\cite{Gang} & ~8.1$\pm$0.41 & 0.00568368 & 0.0034442 & -0.59888 & 1.5 \\
  \hline
  $\textbf{LMC~X-4}$ ~
     $(S_{4})$ & ~1.04$\pm$0.09~\cite{Gang} & ~8.301$\pm$0.2 & 0.00674276 & 0.00429139 & -0.760325 & 1.5 \\
  \hline
  $\textbf{SMC~X-4}$ ~     $(S_{5})$ & ~1.29$\pm$0.05~\cite{Gang} & ~8.831$\pm$0.09 & 0.00728223 & 0.00490204 & -0.950209 & 1.6 \\
  \hline
  $\textbf{4U~1820-30}$ ~
     $(S_{6})$ & ~1.58$\pm$0.06~\cite{Guva} & ~9.1$\pm$0.4 & 0.00873851 & 0.00641177 & -1.25459 & 2 \\
  \hline
  $\textbf{Cen~X-3}$ ~
     $(S_{7})$ & ~1.49$\pm$0.08~\cite{Gang} & ~9.178$\pm$0.13 & 0.00779851 & 0.00551327 & -1.12133 & 1.58 \\
  \hline
  $\textbf{4U~1608-52}$ ~
     $(S_{8})$ & ~1.74$\pm$0.14~\cite{Guv} & ~9.3$\pm$1.0 & 0.00936079 & 0.00720917 & -1.43314 & 3 \\
  \hline
  $\textbf{PSR~J1903+327}$ ~
     $(S_{9})$ & ~1.667$\pm$0.021~\cite{Gang} & ~9.48$\pm$0.03 & 0.00833172 & 0.00617708 & -1.29238 & 2 \\
  \hline
  $\textbf{Vela~X-1}$ ~
     $(S_{10})$ & ~1.77$\pm$0.08~\cite{Gang} & ~9.56$\pm$0.08 & 0.00871723 & 0.00666462 & -1.4058 & 2.5 \\
  \hline
  $\textbf{EXO~1785-248}$ ~
     $(S_{11})$ & ~1.30$\pm$0.2~\cite{Rawls} & ~10.10$\pm$0.44 & 0.00471452 & 0.00302704 & -0.789717 & 2.1 \\
  \hline
\end{tabular}
\label{tab1}
\end{table}
\FloatBarrier

\section{Physical attributes of $f(R,T)$ gravity model}
In this section, we analyze the characteristics of stellar stars in $f(R,T)$ theory of gravity. We investigate the physical behavior of energy density, pressure components, equation of state parameters, surface redshift, energy conditions, stability analysis and equilibrium conditions.
\subsection{Energy Density and Pressure Components}
\begin{figure}[h!]
\begin{tabular}{cccc}
\epsfig{file=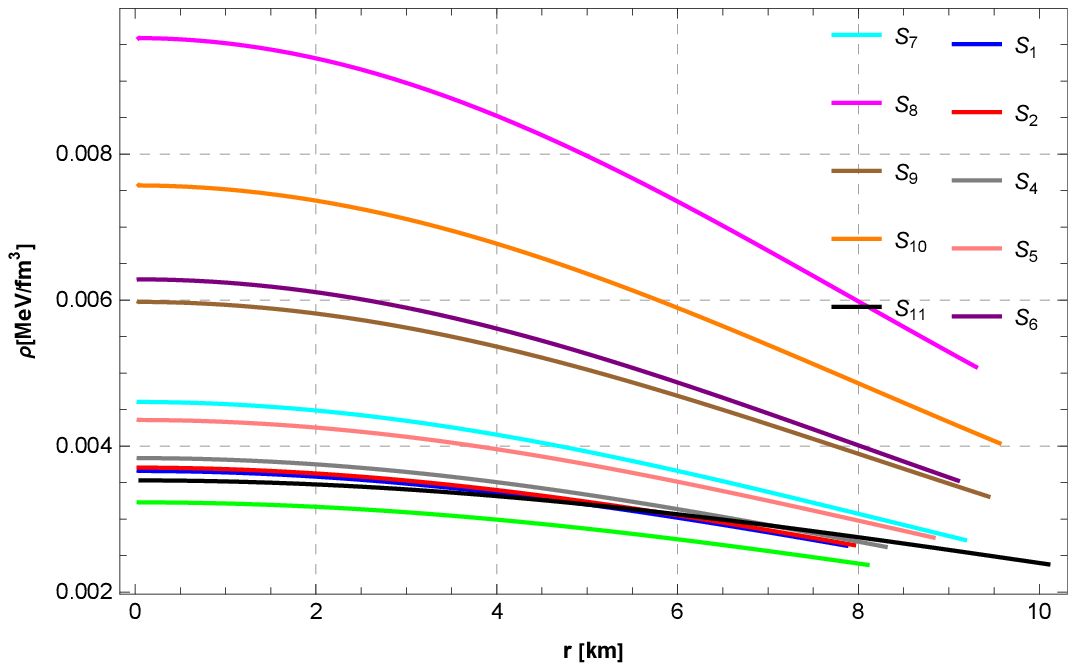,width=0.33\linewidth} &
\epsfig{file=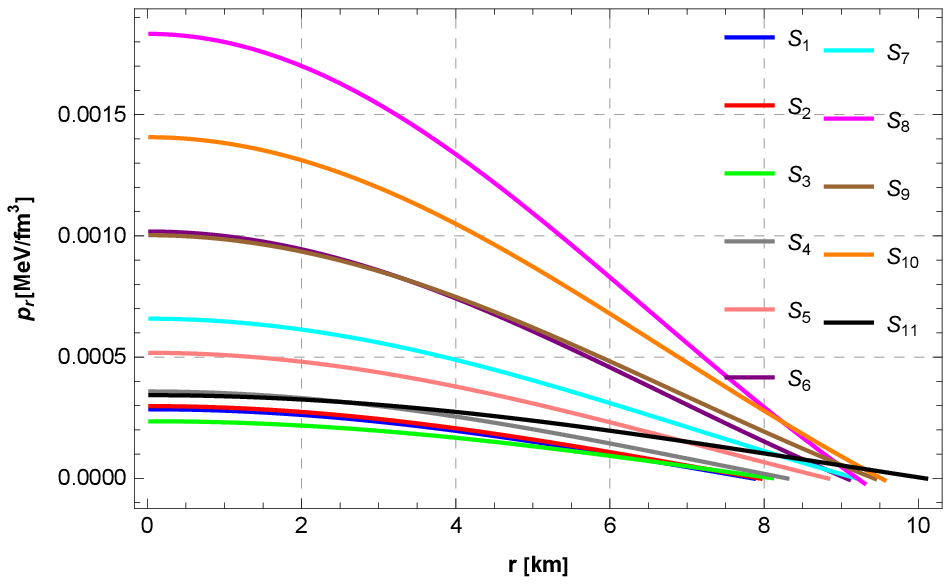,width=0.33\linewidth} &
\epsfig{file=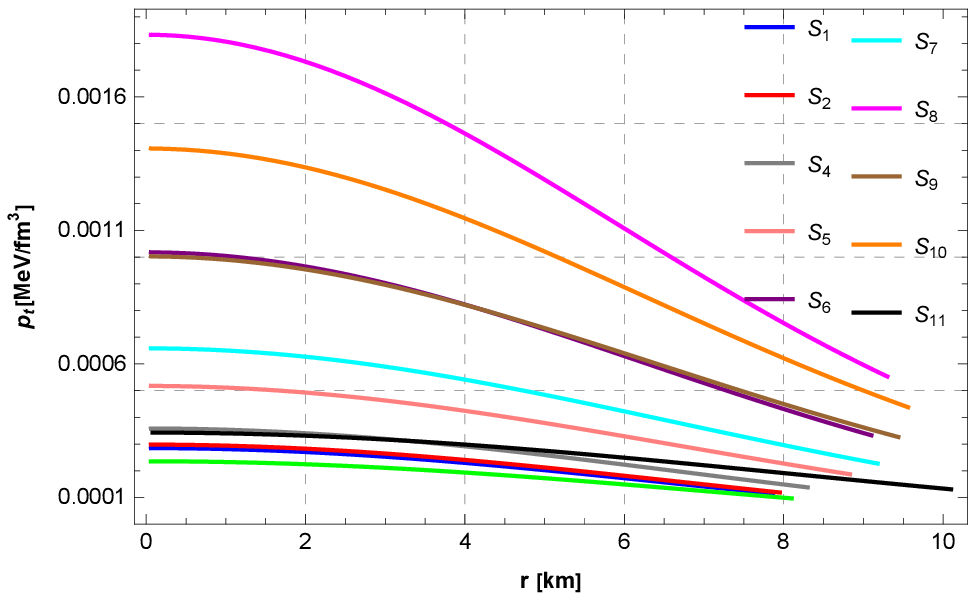,width=0.33\linewidth} &
\end{tabular}
\caption{{Graphical variations of $\rho$, $p_{r}$ and $p_{t}$}.}
\label{Fig:2}
\end{figure}

\begin{figure}[h!]
\begin{tabular}{cccc}
\epsfig{file=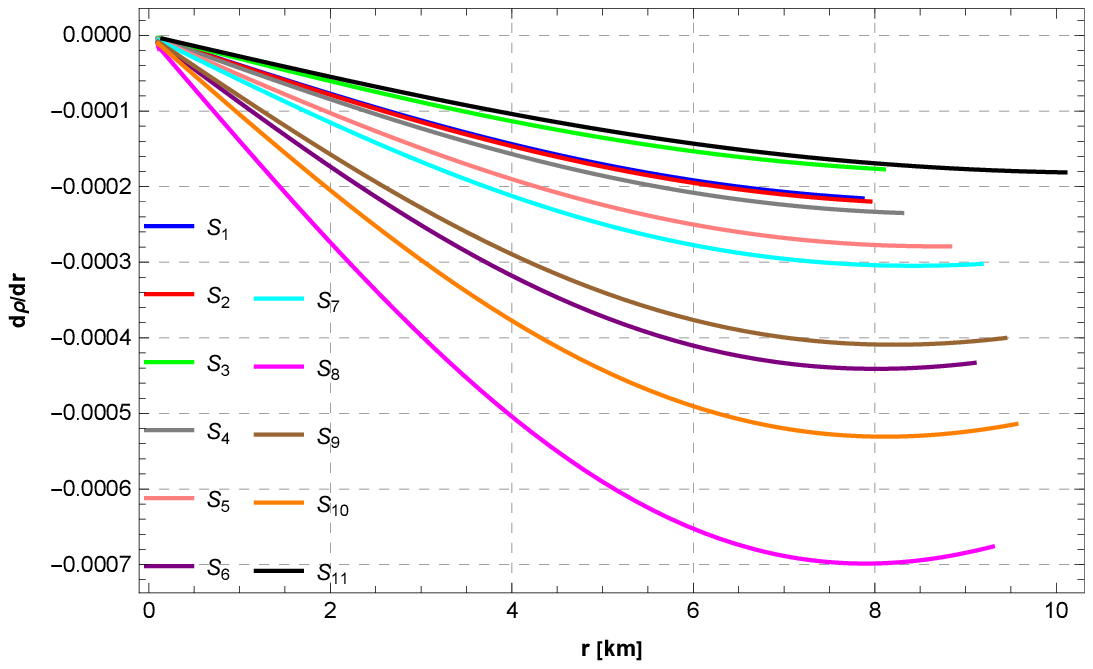,width=0.33\linewidth} &
\epsfig{file=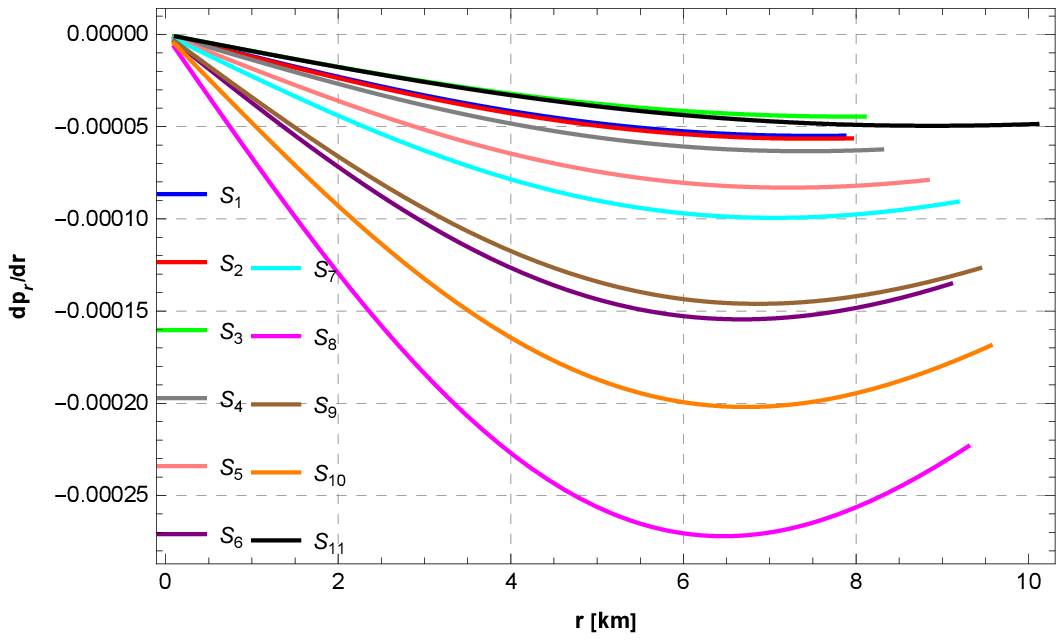,width=0.33\linewidth} &
\epsfig{file=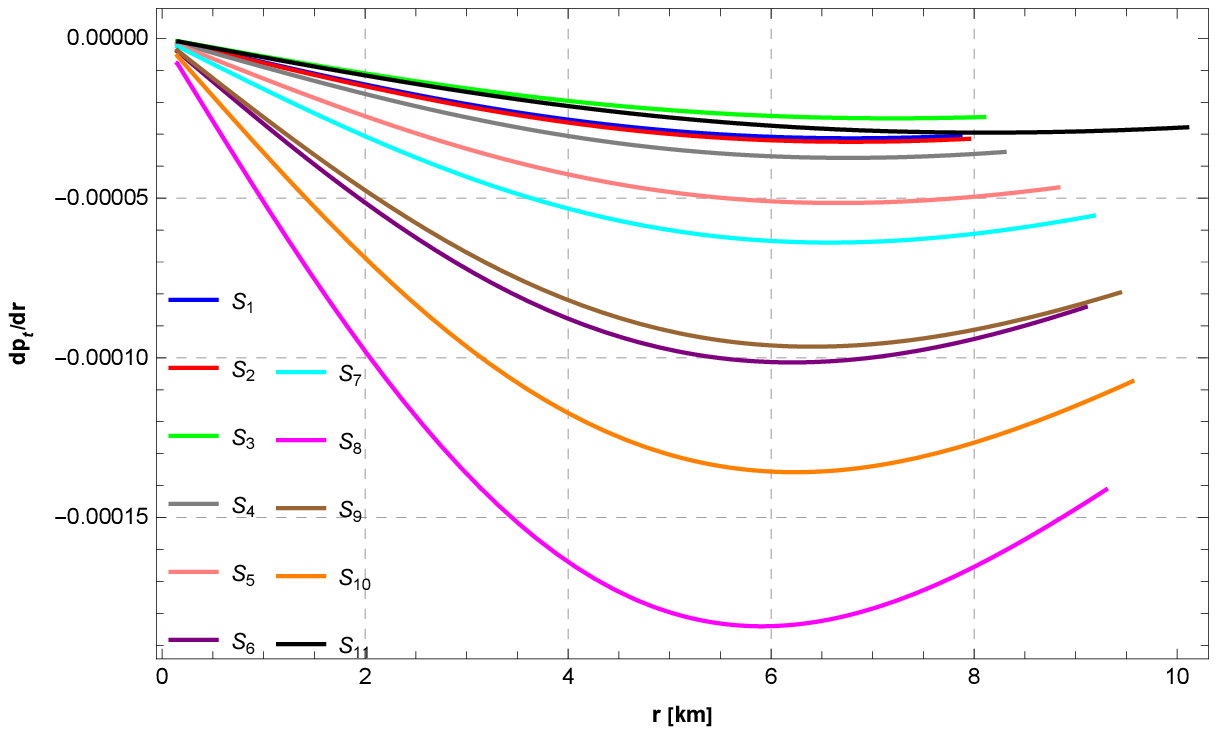,width=0.33\linewidth} &
\end{tabular}
\caption{{Progression of $\frac{dp_{r}}{dr},$ $\frac{dp_{t}}{dr}$ and $\frac{d\rho}{dr}$}.}
\label{Fig:3}
\end{figure}

The graphical representations of energy density, radial pressure and tangential pressure for considered stellar objects are shown in Fig. $\ref{Fig:2}$. The illustrated behavior indicates that the energy density and pressure components achieve their highest value at the core of the stellar stars and then approach zero on the surface boundary. These plots of $\rho,$ $p_r$ and $p_t$ indicate that our system is consistent.
\FloatBarrier
The graphical representation of the rate of change of energy density, radial pressure and tangential pressure i.e., $\frac{d\rho}{dr}$, $\frac{dp_r}{dr}$ and $\frac{dp_t}{dr}$ is zero at the center and becomes negative while moving towards boundary as seen in Fig. $\ref{Fig:3}$. The satisfactory results of these graphs indicate the stability of our proposed $f(R,T)$ model.
\FloatBarrier
\subsection{Tolman-Oppenheimer-Volkoff Equation for $f(R,T)$ Gravity}
\begin{figure}[h!]
\begin{tabular}{cccc}
\epsfig{file=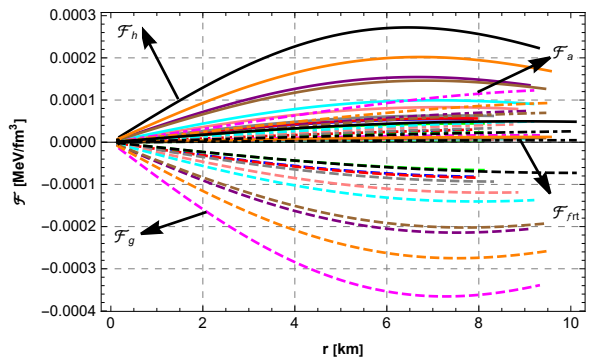,width=0.45\linewidth} &
\end{tabular}
\caption{{Behavior of $\mathcal{F}_{h}$, $\mathcal{F}_{g}$, $\mathcal{F}_{a}$ and $\mathcal{F}_{frt}$}.}
\label{Fig:4}
\end{figure}

The equilibrium state is assumed to be the fundamental attribute in the analysis of compact objects. The geometrical structure of these objects is defined by the Tolman-Oppenheimer-Volkoff (TOV) equation, which helps to explore the equilibrium conditions of stellar structures. The modified form of the generalized $TOV$ equation for the $f(R,T)$ gravity is obtained with the help of Eq. (\ref{3}) and Eq. (\ref{6}) gives
\begin{equation}\label{13}
-\frac{\nu^{'}}{r}(\rho+p_{r})-\frac{dp_{r}}{dr}+\frac{2}{r}(p_{t}-p_{r})+\frac{2\gamma}{3(8\pi-\gamma)}\frac{d}{dr}(3\rho-p_{r}-2 p_{t})=0.
\end{equation}
Firstly, Tolman \cite{Tol} and then Oppenheimer and Volkoff \cite{opp1} demonstrated the equilibrium condition of the stellar structures for the physically stable $f(R,T)$ model. They state that the combination of these forces, namely anisotropic force $\mathcal{F}_{a},$ hydrostatic force $\mathcal{F}_{h},$ gravitational force $\mathcal{F}_{g}$ and
additional expression due to $f(R,T)$ modification $\mathcal{F}_{frt}$ must be zero.
\begin{equation}
\mathcal{F}_{a}+\mathcal{F}_{h}+\mathcal{F}_{g}+\mathcal{F}_{frt}=0,
\end{equation}
where, $ \mathcal{F}_{a}=\frac{2}{r}\Delta,$ $\mathcal{F}_{h}=-\frac{dp_{r}}{dr},$ $\mathcal{F}_{g}=-\frac{\nu^{'}}{r}(\rho+p_{r})$ and $\mathcal{F}_{frt}=\frac{2\gamma}{3(8\pi-\gamma)}\frac{d}{dr}(3\rho-p_{r}-2 p_{t})$. The graphical representation of all forces can be seen in  Fig. $\ref{Fig:4},$ which indicates that our system fulfills all equilibrium conditions and exhibits the viability of our chosen model.
\FloatBarrier
\subsection{Energy Conditions}
\begin{figure}[h!]
\begin{tabular}{cccc}
\epsfig{file=pr2.eps,width=0.33\linewidth} &
\epsfig{file=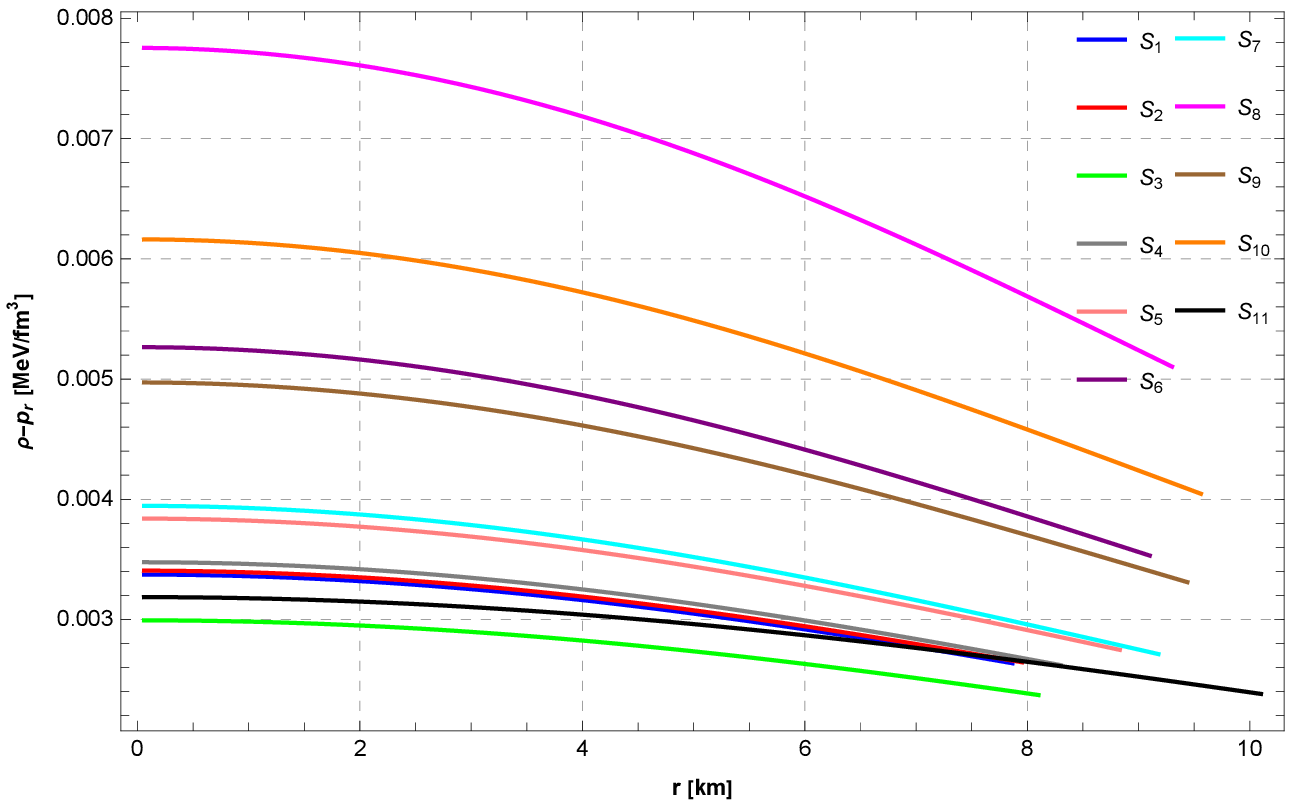,width=0.33\linewidth} &
\epsfig{file=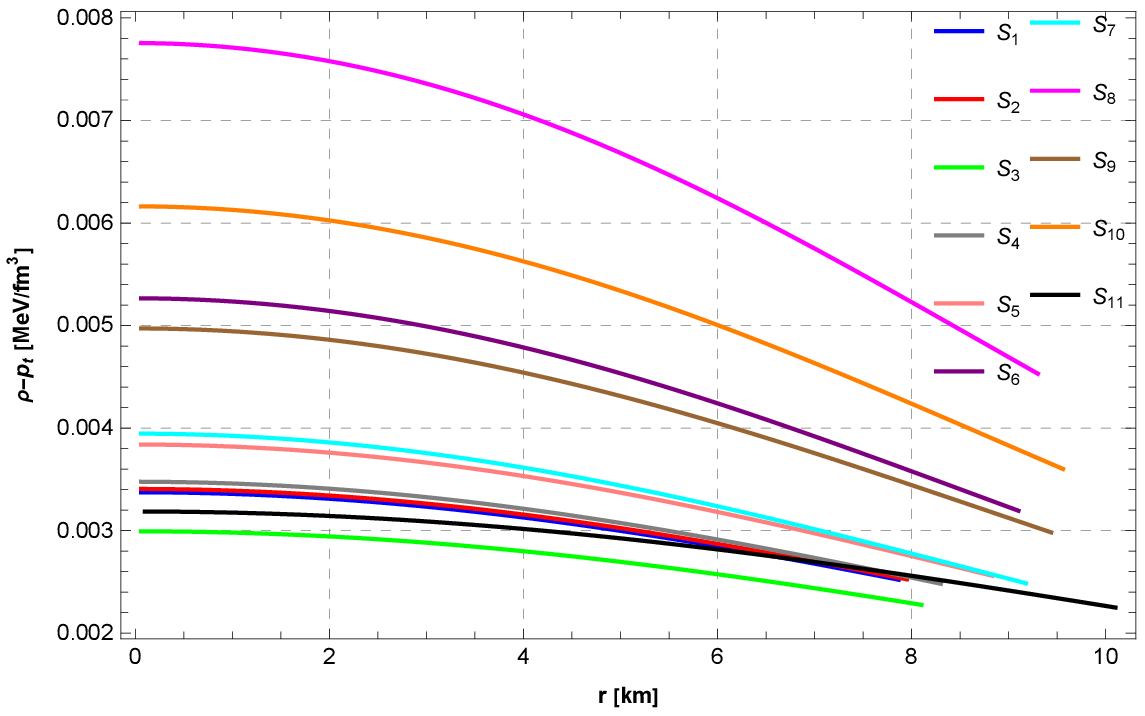,width=0.33\linewidth} &\\
\epsfig{file=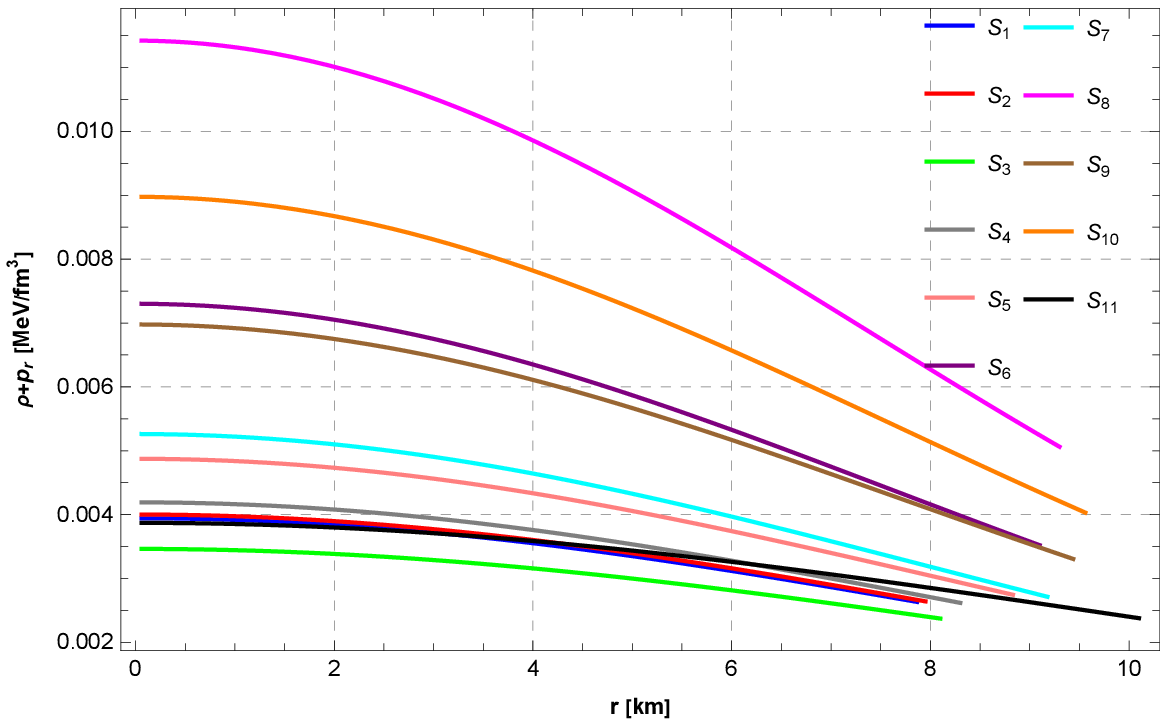,width=0.33\linewidth} &
\epsfig{file=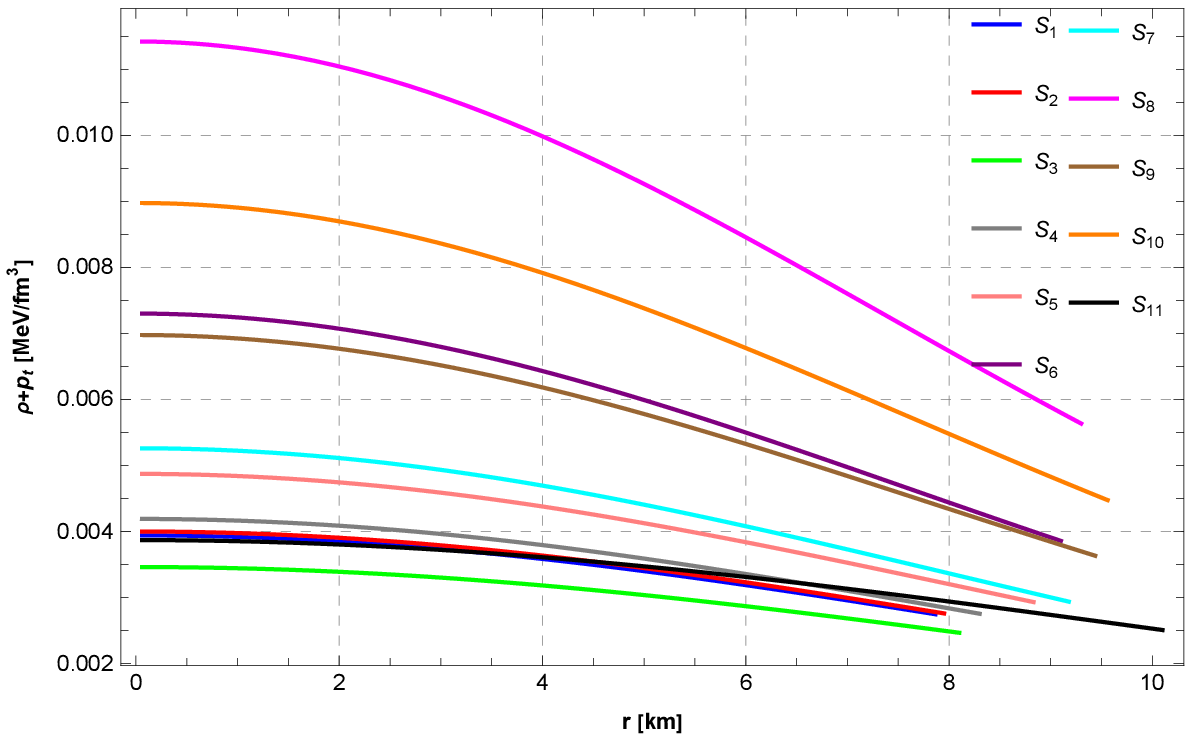,width=0.33\linewidth} &
\epsfig{file=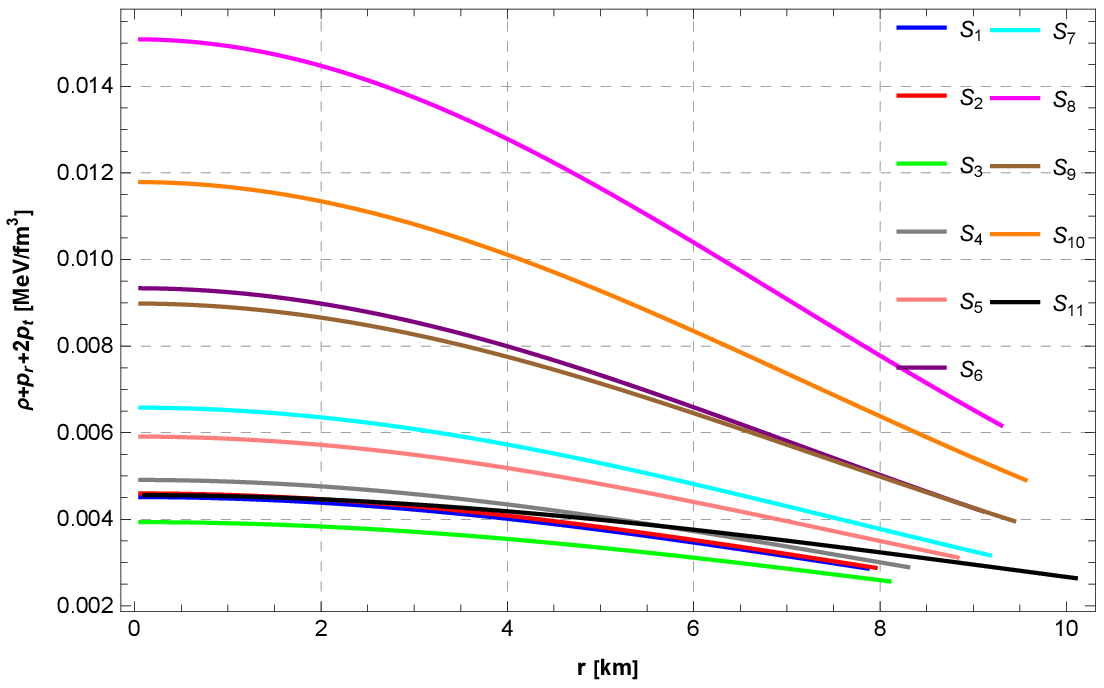,width=0.33\linewidth} &
\end{tabular}
\caption{{Evolution of energy bonds}.}
\label{Fig:5}
\end{figure}

The discussion about the existence of an anisotropic configuration is about its physical attributes, such as its energy conditions. Moreover, these conditions are essential for determining the exotic behavior of matter within the compact structure model. Energy conditions are organized as dominant energy conditions $(DEC),$ Null energy conditions $(NEC),$ Weak energy conditions $(WEC),$ and Strong energy conditions $(SEC)$,
\begin{equation*}
DEC:\rho- p_{r} > 0,\rho- p_{t} > 0,~~~~~~~~ NEC:  \rho+p_{r}\geq 0,\rho+p_{t}\geq 0,
\end{equation*}
\begin{equation*}
WEC:\rho \geq0,\rho+p_{r} \geq 0,\rho+p_{t} \geq 0,~~~~~~~~SEC: \rho+p_{r}\geq 0,\rho+p_{t}\geq 0, \rho+p_{r}+2p_{t}\geq 0.
\end{equation*}
The graphical analysis of all the energy conditions such as $NEC,$ $WEC,$ $SEC$ and $DEC$ are satisfied and can be seen in Fig. $\ref{Fig:5}$, which justifies that our star is viable.
\FloatBarrier
\subsection{Equation of State $(EoS)$ Parameters}
\begin{figure}[h!]
\begin{tabular}{cccc}
\epsfig{file=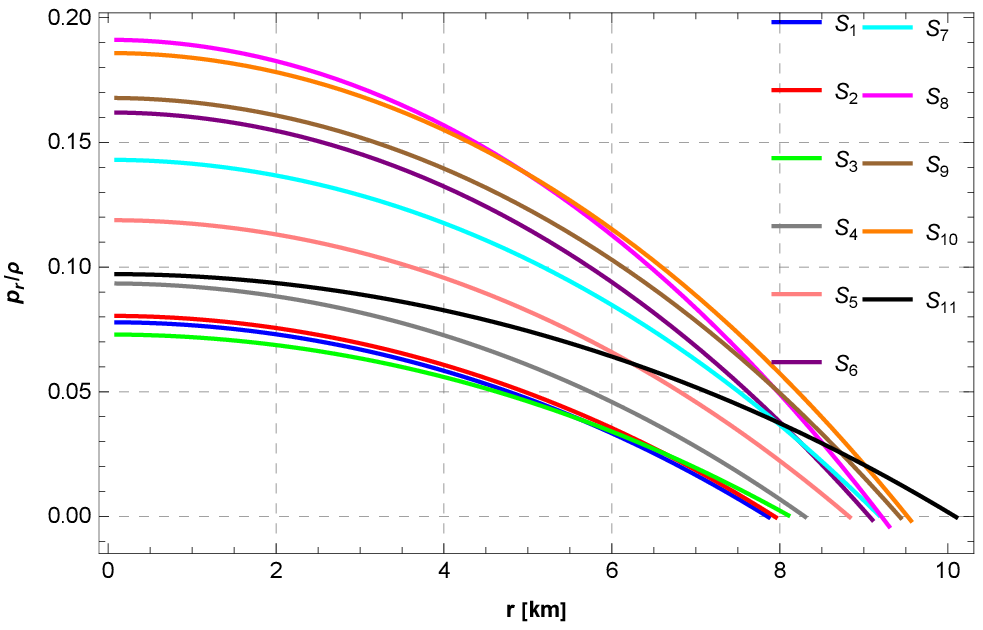,width=0.33\linewidth} &
\epsfig{file=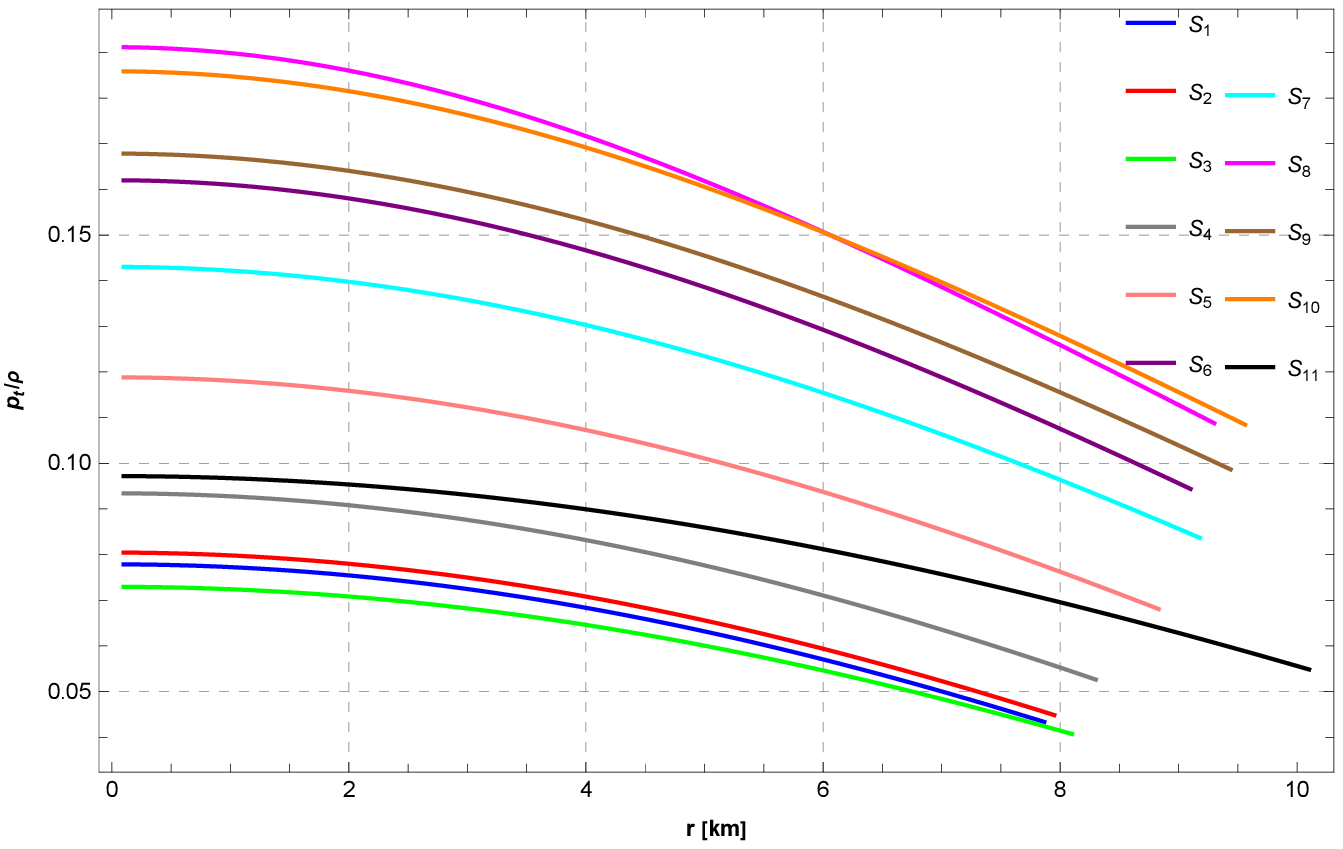,width=0.33\linewidth} &
\end{tabular}
\caption{{Behavior of $w_r$ and $w_{t}$}.}
\label{Fig:6}
\end{figure}

In literature, there are different types of equation of state parameters. For our current work, we consider two types of $EoS$ parameters namely radial $EoS$ parameter and transverse $EoS$ parameter, defined as
\begin{equation}
   w_{r} = \frac{p_{r}}{\rho},~~~~~~~~w_{t} =  \frac{p_{t}}{\rho}.
\end{equation}
The graphical representation of these two parameter can be seen in Fig. $\ref{Fig:6},$ which clearly shows that $0<w_{r},~w_{t}<1.$ It suggests that the nature of our system is well-behaved.
\FloatBarrier
\subsection{Anisotropy}
\begin{figure}[h!]
\begin{tabular}{cccc}
\epsfig{file=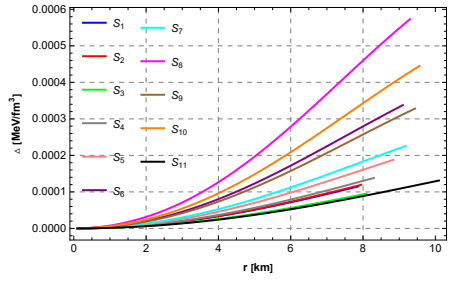,width=0.33\linewidth} &
\end{tabular}
\caption{{Variations of $\Delta$}.}
\label{Fig:7}
\end{figure}

We investigate the internal structure of relativistic compact stars with the help of anisotropic factor defined as $\Delta= p_t-p_r$ \cite{Ani}. The anisotropic factor is the inward direction when $p_t < p_r,$ indicating $\Delta <0,$ but in the case of $p_t > p_r,$ the direction of the anisotropy is outward, which implies $\Delta > 0$ \cite{Ani1}. The anisotropy for a chosen $f(R,T)$ gravity model demonstrates a repulsive behavior as seen in Fig. $\ref{Fig:7}.$
\FloatBarrier
\subsection{Mass-Radius Relationship, Compactness factor and Surface Redshift Evolution}
\begin{figure}[h!]
\begin{tabular}{cccc}
\epsfig{file=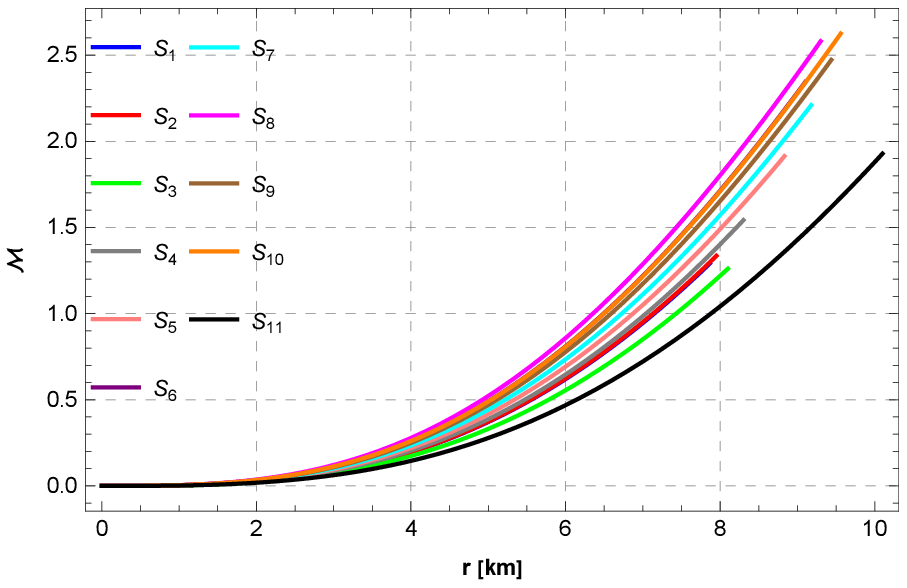,width=0.33\linewidth} &
\epsfig{file=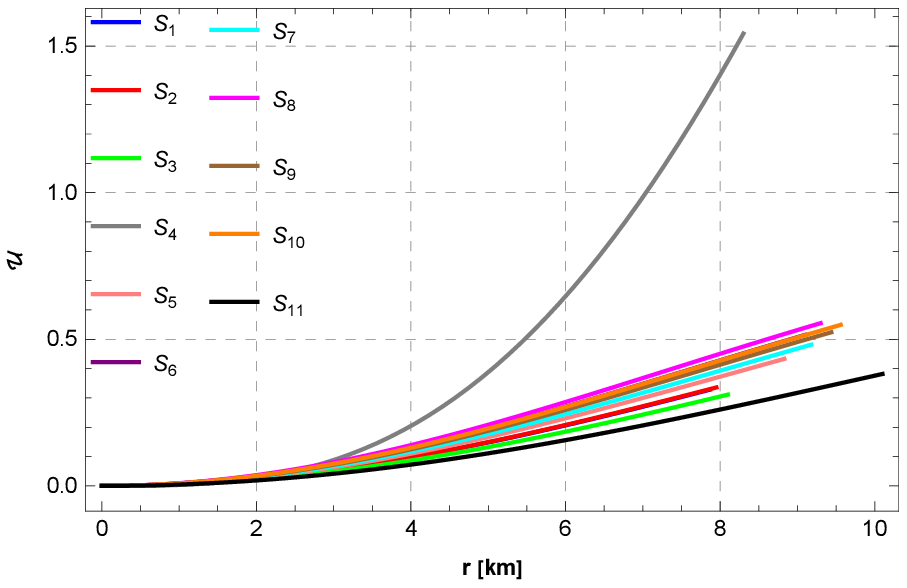,width=0.33\linewidth} &
\epsfig{file=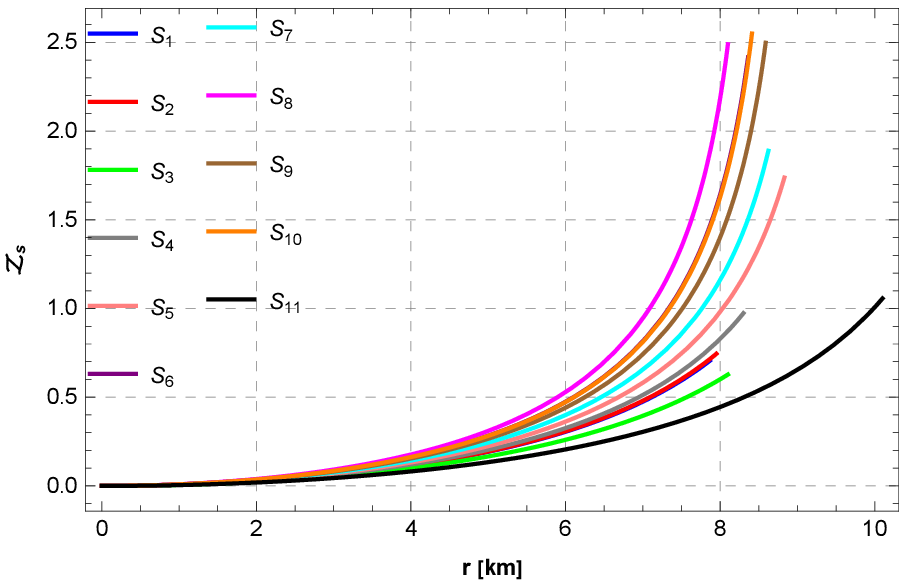,width=0.33\linewidth} &
\end{tabular}
\caption{{Behavior of $\mathcal{M}(r) (left panel),$ $\mathcal{U}(r) (center panel),$  and $\text{$\mathcal{Z}$s} (right panel)$}.}
\label{Fig:8}
\end{figure}

By utilizing metric potential $g_{rr}^{-}$ = $g_{rr}^{+}$, the mass function \cite{Bhar12} is derived by
\begin{equation}
\mathcal{M}(r)=\frac{r(e^{Ar^{2}}-1)}{2e^{Ar^{2}}}.
\end{equation}
It can be seen from the left panel of Fig. $\ref{Fig:8}$ that the graphical representation of mass function is increasing as we move towards boundary, which means that mass is uniform at the center, i.e. $\mathcal{M}(r) \rightarrow 0$ as $r\rightarrow 0.$ Furthermore, the compactness parameter $\mathcal{U}(r)$ \cite{MakHar} and surface redshift $\text{$\mathcal{Z}$s}$ \cite{Boh} have the following mathematical configuration:
\begin{equation}
\mathcal{U}(r)=\frac{2\mathcal{M}(r)}{r} = 1-\frac{1}{e^{Ar^{2}}},
\end{equation}
\begin{equation}
  \text{$\mathcal{Z}$s} = \frac{1}{\sqrt{1-2\nu}}-1.
\end{equation}
The physical illustrations of compactness parameter and surface redshift is positive as well as increasing nature similar to mass function, shown in the middle and right panel of  Fig. $\ref{Fig:8}$.
\FloatBarrier
\subsection{Causality Condition and Stability Analysis}
\begin{figure}[h!]
\begin{tabular}{cccc}
\epsfig{file=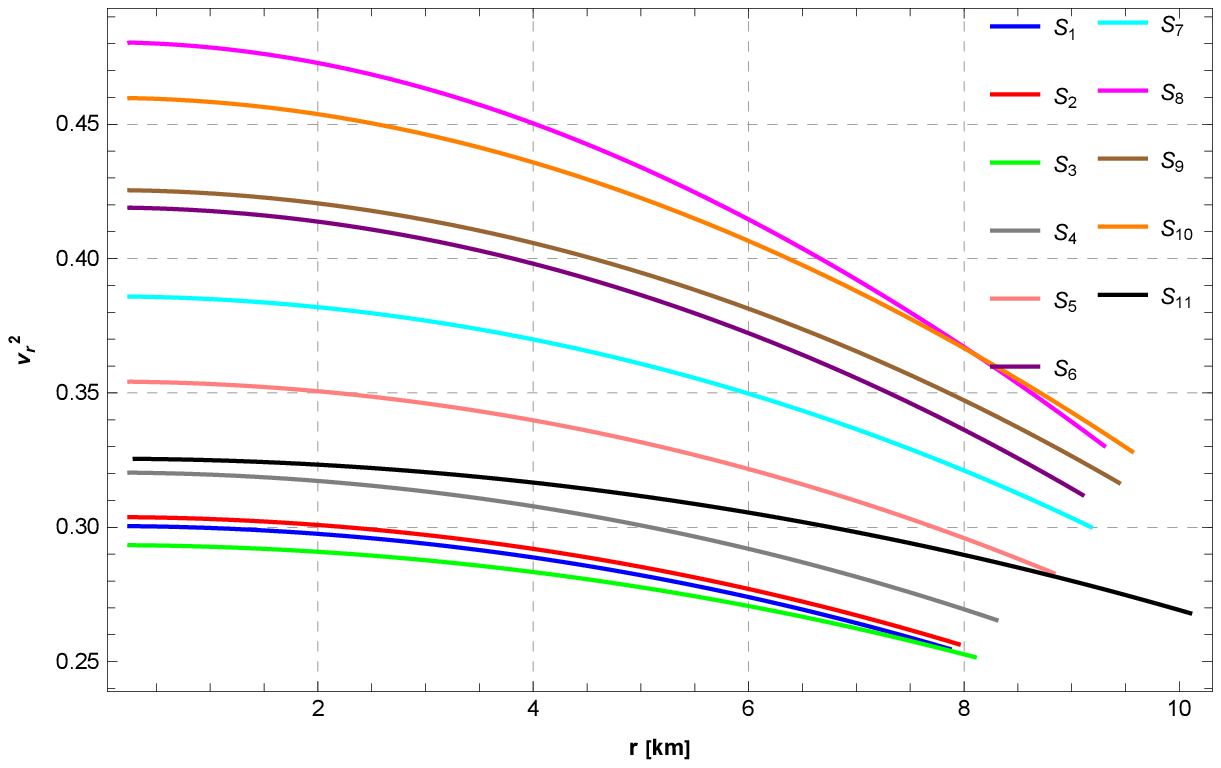,width=0.33\linewidth} &
\epsfig{file=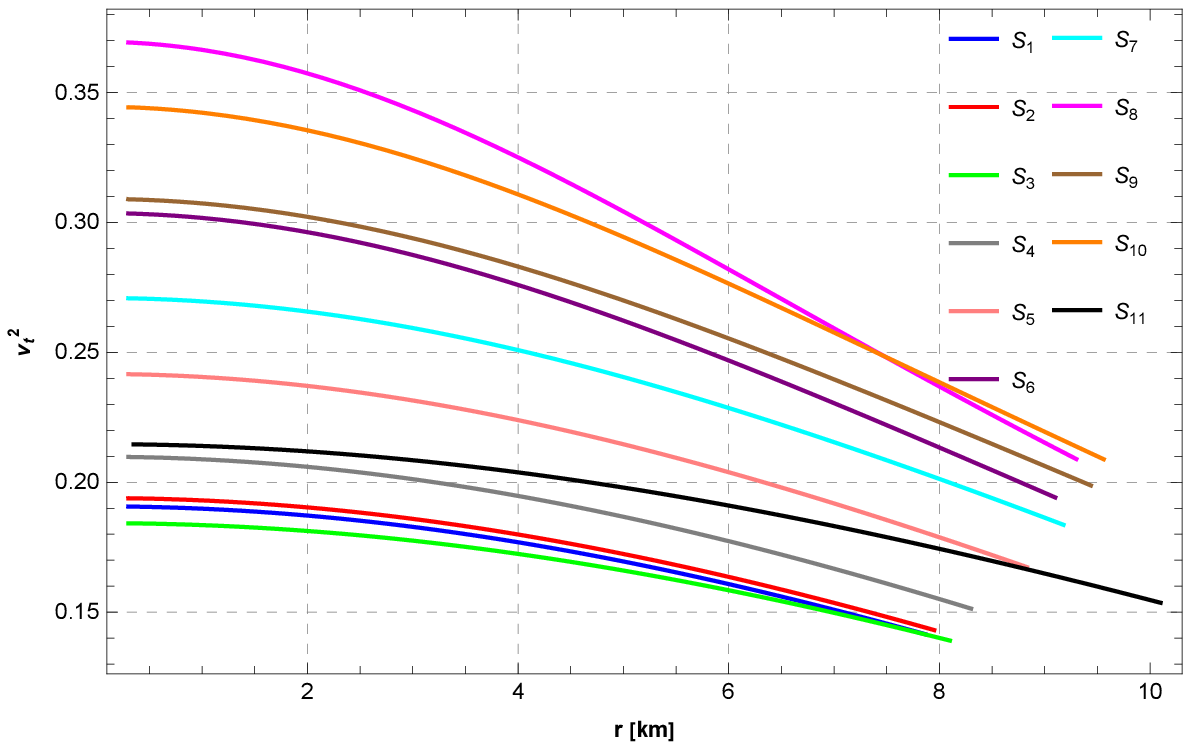,width=0.33\linewidth} &
\epsfig{file=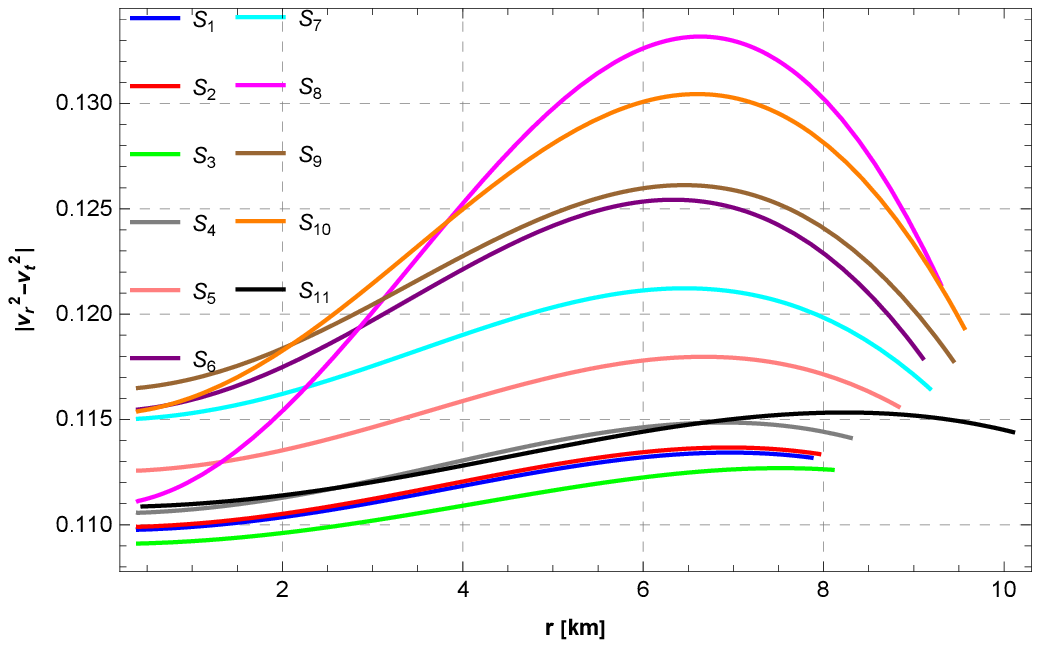,width=0.33\linewidth} &
\end{tabular}
\caption{{Variations of $v_{r}^{2}$, $v_{t}^{2}$ and $\mid v_{t}^{2}-v_{r}^{2}\mid$}.}
\label{Fig:9}
\end{figure}

Here, we discuss the stability analysis in $f(R,T)$ theory of gravity. The causality condition plays a significant role for the discussion of realistic stars. The sound speed of the pressure waves must be less than the speed of the light waves when $f(R,T)$ model exhibits an internal configuration. When it comes to anisotropic fluids, the causality condition of the pressure waves arises in radial and transverse directions. Therefore, both components must be restricted to the speed of light to achieve a physically consistent $f(R,T)$ model. The causality condition is based on the following expression
\begin{equation}
  {v_{r}}^{2} = \frac{dp_{r}}{d\rho},~~~~~~~~{v_{t}}^{2} = \frac{dp_{t}}{d\rho},
\end{equation}
where, ${v_{r}}^{2}$ stands for radial sound speed and ${v_{t}}^{2}$ represents the transverse sound speed. For a physically stable celestial structure, the radial and transverse sound speeds must be the range $[0,1]$. We use the Herrera approach \cite{Herr} $(i.e. 0\leq v_{r}^{2}, v_{t}^{2}\leq1)$ to explore the stability of our $f(R,T)$ model. Also, to ensure the consistent behaviour of our proposed model, we considered the Andreasson condition \cite{Andr} i.e. $|{v_{r}}^{2}-{v_{t}}^{2}|\leq1.$ It is significant to indicate that $f(R,T)$ model under analysis is fully balanced, as seen in Fig. $\ref{Fig:9}$.
\FloatBarrier
\section{Conclusion}
Various modified theories are considered a fundamental aspect of astrophysics to investigate the complex nature of the stellar configuration. Among all modified theories, cosmologists are especially interested in $f(R,T)$ gravity because of its unique combination of curvature and matter. Also, $f(R,T)$ demonstrated several remarkable results in the area of thermodynamics, cosmology, and the astrophysics of relativistic stars. Several promising outcomes indicate that $f(R,T)$ gravity has already been probed in astrophysics. Our goal is to study the precise nature of stellar stars while regarding the realistic $f(R,T)$ gravity model. To accomplish our objective, we presume the metric potentials using the Krori-Barua model, $\nu(r) = Br^{2} + C$ and $\lambda(r) = Ar^{2},$ where $A,$ $B,$ and $C$ are unknowns. Further, to evaluate these constants compare the internal geometry to the exterior Schwarzschild sphere. The graphical representations of several compact stars ensure the consistency of our results. The main focus is to achieve an appropriate family of $f(R,T)$ gravity solutions in anisotropic distribution. The prominent findings are detailed below:
\begin{itemize}
\item It is common practice to describe the nature of spacetime using metric potentials. Fig. $\ref{Fig:1}$ indicates that the metric potentials $g_{tt} = e^{\nu}$ and $g_{rr} = e^{\lambda}$ have positive, singularity-free graphs that satisfy all conditions i.e., $e^{\nu(r=0)} = e^{C}$ and $e^{\lambda(r=0)} = 1$. The evolution of both potentials shows satisfactory results since these graphs increase monotonically, indicating the validity of the $f(R,T)$ gravity model.

\item Fig. $\ref{Fig:2}$ demonstrates the behavior of the density and pressure components for the $f(R,T)$ gravity model. It is also clear that plots of $\rho,$ $p_r,$ and $p_t$ obtain their maximum values at the center and exhibit a decreasing response toward the boundary, which consistency the stability of our model.

\item Fig. $\ref{Fig:3}$ shows the gradient of the pressure components and energy density. These graphs show negative illustrations, indicating that our outcomes are reliable.

\item All forces $\mathcal{F}_{h},$ $\mathcal{F}_{g},$ $\mathcal{F}_{a},$ and $\mathcal{F}_{frt}$ are steady and show balancing behavior, as seen in Fig. $\ref{Fig:4}$.

\item Fig. $\ref{Fig:5}$ provides the energy bonds for the proposed $f(R,T)$ model. It is important to highlight that our model satisfies all energy conditions.

\item The physical illustration of $EoS$ parameters satisfy the conditions $0 \leq w_{r}$ and $w_{t} \leq 1$ for the compact objects to behave precisely. Fig. $\ref{Fig:6}$ exhibits that the nature of both $EoS$ ratios are steady.

\item The graphical representation of anisotropy in Fig. $\ref{Fig:7}$ illustrates the stable and repulsive nature of the stellar star.

\item From Fig. $\ref{Fig:8}$, it is easy to notice that the mass function, compactness factor, and surface redshift increase monotonically.

\item The parameters of sound velocity $v_{r}^{2}$ and $v_{t}^{2}$ lie between $[0, 1]$ for the proposed $f(R,T)$ model. The stability constraints for our model are stable, as shown in Fig. $\ref{Fig:9}$.
\end{itemize}
Modified $f(R,T)$ gravity plays an alluring role in the study of stellar structures. Also, it is significant to note that our obtained outcomes are much similar to the results discussed by Shamir et al. \cite{Naz1} in $f(R)$ gravity. Thus we conclude that our $f(R,T)$ model in the frame of anisotropic fluid is stable and consistent, as all physical characteristics of compact objects obey physically obtained patterns.

\end{document}